\documentclass{Interspeech}
\usepackage{multirow} % 处理多行合并
\usepackage{caption}  % 自定义标题
\usepackage{lipsum}   % 生成示例文本
\usepackage{booktabs}
\usepackage{amsmath}
\usepackage{array}
\usepackage{float}
\usepackage{comment}
\usepackage{hyperref}

\interspeechcameraready

\title{Overlap-Adaptive Hybrid Speaker Diarization and ASR-Aware Observation Addition for MISP 2025 Challenge}

\author[affiliation={1}{*}]{Shangkun}{Huang}
\author[affiliation={1}{*}]{Yuxuan}{Du}
\author[affiliation={1}]{Jingwen}{Yang}
\author[affiliation={1}]{Dejun}{Zhang}
\author[affiliation={1}]{Xupeng}{Jia}
\author[affiliation={1}]{Jing}{Deng}
\author[affiliation={2,3}]{Jintao}{Kang}
\author[affiliation={1}]{Rong}{Zheng}

\affiliation{}{Beijing Fosafer Information Technology Co., Ltd.}{China}
\affiliation{}{Institute of Forensic Science, Ministry of Public Security}{China}
\affiliation{}{The Institute of Linguistics, Chinese Academy of Social Sciences}{China}
\email{\{huangshangkun, duyuxuan\}@fosafer.com}


\usepackage{etoolbox} % 引入 etoolbox 用于修改命令

\makeatletter
\renewcommand\NoHyper{}
\makeatother

\usepackage[dvipsnames]{xcolor} % 引入xcolor宏包，dvipsnames选项提供了更多预设颜色

\definecolor{myblue}{RGB}{0, 0, 255}      % 定义一种深邃的蓝色
\definecolor{mygreen}{RGB}{46, 139, 87}    % 定义一种海绿色
\definecolor{myred}{RGB}{192, 25, 50}      % 定义一种不刺眼的红色

\hypersetup{
    colorlinks=true,
    linkcolor=myblue,
    citecolor=blue,
    urlcolor=blue,
    pdfborder={0 0 0},
}

\begin{document}

\maketitle
\renewcommand{\thefootnote}{\fnsymbol{footnote}}
\footnotetext[1]{Equal contribution.}
\renewcommand{\thefootnote}{\arabic{footnote}}

\begin{abstract}
This paper presents the system developed to address the MISP 2025 Challenge. For the diarization system, we proposed a hybrid approach combining a WavLM end-to-end segmentation method with a traditional multi-module clustering technique to adaptively select the appropriate model for handling varying degrees of overlapping speech. For the automatic speech recognition (ASR) system, we proposed an ASR-aware observation addition method that compensates for the performance limitations of Guided Source Separation (GSS) under low signal-to-noise ratio conditions. Finally, we integrated the speaker diarization and ASR systems in a cascaded architecture to address Track 3. Our system achieved character error rates (CER) of 9.48\% on Track 2 and concatenated minimum permutation character error rate (cpCER) of 11.56\% on Track 3, ultimately securing first place in both tracks and thereby demonstrating the effectiveness of the proposed methods in real-world meeting scenarios.
    
\end{abstract}

\begingroup
  \renewcommand\thefootnote{*}% 临时把脚注符号改成 *
  % \footnotetext{The authors contribute equally.}
  \addtocounter{footnote}{-1}% 避免与后文普通数字脚注冲突
\endgroup

\begin{figure*}[htb]
    \centering
    \includegraphics[width=1\textwidth]{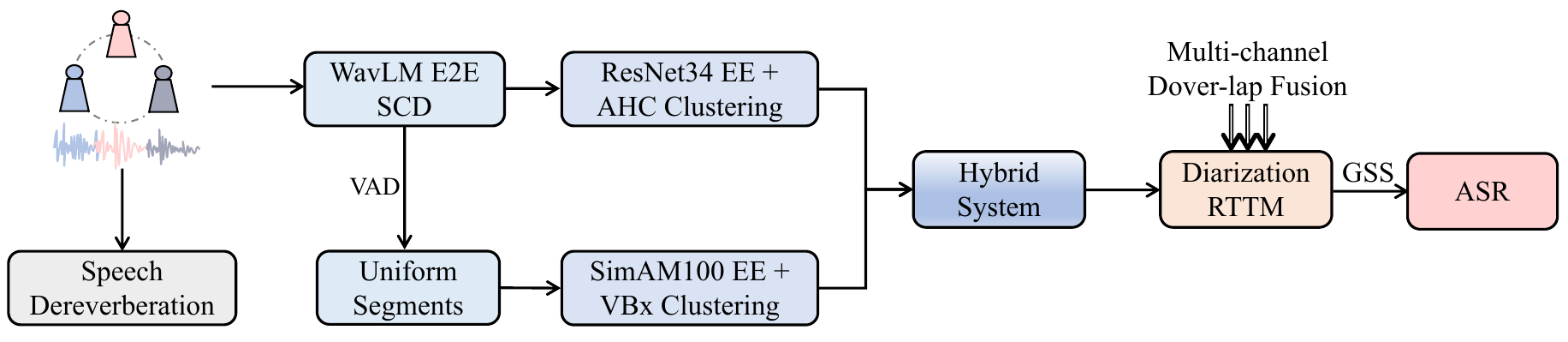} % 调整图片大小，width=\textwidth 使其占满整行
    \caption{The overall workflow of the combined diarization and ASR system submitted for MISP2025 challenge.}
    \label{fig:dudu1} % 给图片加标签，方便引用
\end{figure*}

\section{Introduction}

The MISP 2025 Challenge\footnote[1]{https://mispchallenge.github.io/mispchallenge2025/index.html} aims to advance the application and development of multimodal speech processing technologies in meeting contexts\cite{misp2025zongjie}. This challenge provides an open-source, state-of-the-art, large-scale multimodal meeting dataset that encompasses meetings of various sizes and rich topic transitions\cite{2025misp}. The competition comprises three tracks: audio-visual speaker diarization (AVSD), audio-visual speech recognition (AVSR), and audio-visual diarization and recognition (AVDR). The AVSD track addresses the "who spoke when" problem. The AVSR track involves effectively handling overlapping speech segments and recognizing content from multiple speakers. The AVDR track integrates the first two tasks, aiming to resolve the "who spoke what when" problem.

With advancements in speech processing, AVDR techniques have become increasingly essential in complex acoustic environments. End-to-end deep learning models, such as Target Speaker Voice Activity Detection (TS-VAD), have emerged as key research topics in diarization field, effectively handling overlapping speech through speaker embedding and improving performance in challenging conditions\cite{introduction1,introduction2}. Audio-visual fusion methods have also been widely applied, using video data to assist in speech activity detection, particularly in noisy environments, compensating for limitations in speech data\cite{introduction3,introduction4,9-misp-3wang2023multimodal}. In the MISP 2022 Challenge, several systems improved multi-speaker performance by jointly training audio and video models, thereby enhancing speech recognition accuracy\cite{9-misp-3wang2023multimodal,introduction6}. Additionally, some techniques such as microphone array-based GSS and self-supervised learning have been employed to refine audio quality and enhance model robustness\cite{introduction7,introduction8}. The integration of these methods continues to advance AVDR, showcasing significant potential in complex scenarios.

This paper presents a technical report on our submissions to the three tracks of the MISP2025 Challenge. For the AVSD task, we propose a hybrid speaker diarization system that combines traditional multi-module methods with end-to-end segmentation approaches. The end-to-end approach leverages the self-supervised learning-based WavLM model to minimize reliance on large-scale datasets, whereas the traditional method employs dereverberation techniques and Bayesian HMM clustering of x-vector sequences (VBx). By dynamically selecting methods suited to varying levels of overlapping speech, our system achieved a DER of 8.88\% in the Track 1 evaluation set. For the AVSR system, we propose the ASR-Aware OA framework to alleviate the performance bottleneck of the GSS algorithm in low-SNR environments. This framework innovatively fuses noisy speech, Mossformer2-separated speech, and GSS-separated speech, and introduces a sentence-level bridging module supervised by ASR to optimize OA coefficients, with CER as the training metric. Our system final achieved a CER of 9.48\% in Track 2. The AVDR system combines methods from the first two tasks, ultimately achieving a cpCER of 11.56\%.

\section{System Description}

\autoref{fig:dudu1} illustrates the overall workflow of the combined diarization and ASR system submitted for MISP2025 challenge.

\subsection{Speaker Diarization}

In the diarization task, we employ a hybrid system that integrates traditional multi-module methods with end-to-end segmentation approaches. Specifically, for the end-to-end segmentation method, we leverage the self-supervised learning-based speaker diarization approach proposed by \cite{wavlm}. This method replaces the former segmentation module, which is based on the SincNet+LSTM architecture \cite{pyannote3}, with a novel architecture built on WavLM and Conformer. We further improve diarization performance by replacing the WavLM-Base+ used in \cite{wavlm} with WavLM-Large. Once the initial segmentation results are obtained from the model, we extract speaker embeddings using the ResNet-based embedding extractors and apply hierarchical agglomerative clustering (AHC) to obtain the final clustering results. Additionally, the segmentation results from this model are also used to obtain the voice activity detection (VAD) results in the traditional multi-module system and to calculate the proportion of overlapping speech duration. For the traditional multi-module approach, we employ the segmentation results to obtain the initial VAD results. Additionally, we use an MP-SENet-based enhancement model for front-end dereverberation processing\cite{mp}.  Although the speech enhancement model effectively suppresses noise and reverberation in meetings, it may introduce artifacts that affect the similarity of speaker embeddings. To address this issue, we apply a scaling and mixing strategy to the original and enhanced speech signals to generate the final input audio. Subsequently, we use SimAM-ResNet100 \cite{simam}, provided by \cite{wespeaker}, as the embedding extractor. For clustering, we adopt a diarization approach based on VBx \cite{vbx}.

After obtaining the results from these methods, we use the proportion of overlapping speech to determine which method to apply for each meeting scenario. Specifically, the VBx method performs more effectively in scenarios with a low proportion of overlapping speech, whereas the end-to-end approach is better suited for handling the assignment of speakers in overlapping speech. Therefore, based on the development set results, we assign meeting scenarios with less than 1$\%$ overlapping speech to the traditional method, and assign the remaining scenarios to the segmentation method. All single-channel tests are conducted on channel 1, and we finally use Dover-lap \cite{dover} to fuse the results from eight channels.

\subsection{Speech Recognition}
\label{subsec: asr}

In the ASR field, low SNR environments can limit the performance of the Guided Source Separation algorithm\cite{11-gss-raj2022gpu}, making it challenging to balance noise suppression and speech fidelity, thereby affecting recognition accuracy\cite{14-han2024audio}. Although neural network-based compensation solutions can mitigate signal loss, they are limited by the mismatch between the objective functions of the frontend separation model and the backend recognition module\cite{1-huang2024fosafer,2-huang2024enhanced}. As a result, the enhanced speech still contains time-frequency artifacts, reducing the robustness of ASR systems. Observation Addition (OA)\cite{5-oa-1cui2025reducing} effectively mitigates these artifacts by adding the original noisy speech to the separated speech with a specific coefficient before passing it into the ASR module for recognition.

\begin{figure*}[htb]
    \centering
    \includegraphics[width=0.7\textwidth]{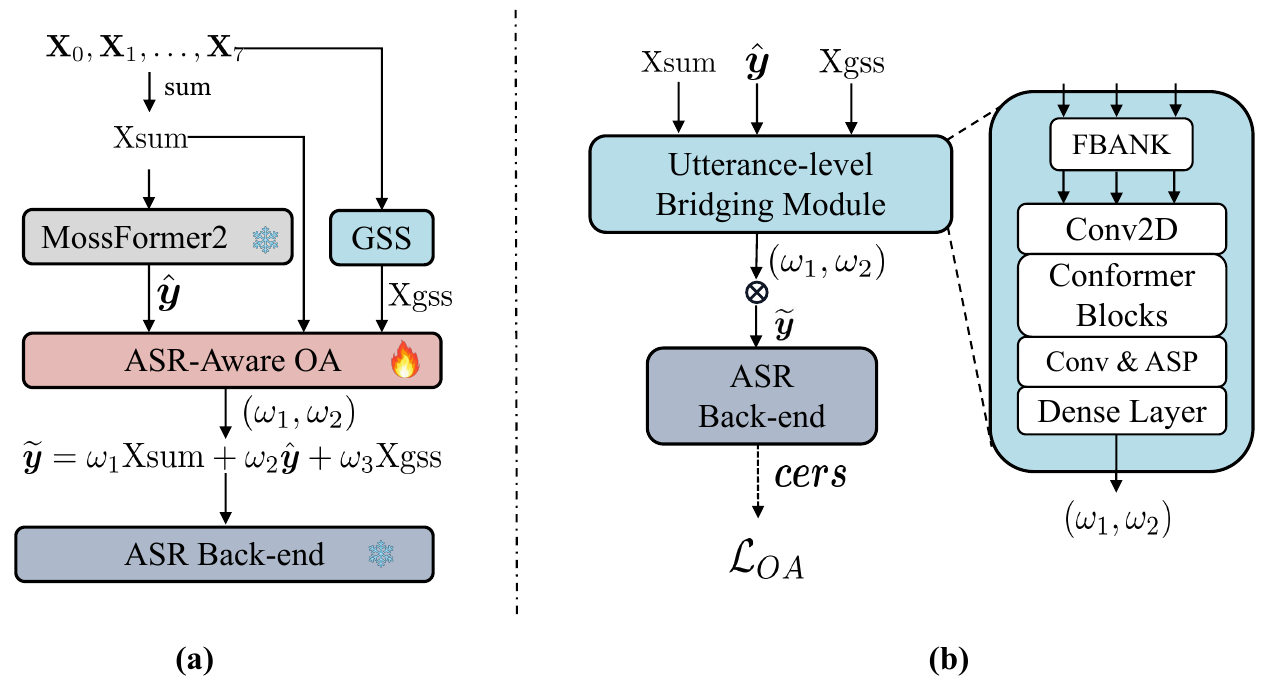} % 调整图片大小，width=\textwidth 使其占满整行
    \vspace{-0.5em} % 插入 1em 的垂直空白，可以根据需要调整
    \caption{Our proposed ASR-aware observation-adding system (a) Overall framework (b)Bridging module structure }
    \label{fig:kunkun1} % 给图片加标签，方便引用
    
\end{figure*}

To overcome the limitations of GSS, inspired by\cite{5-oa-1cui2025reducing}, we propose an enhanced OA method that combines weighted noisy speech, speech separated (SS) by Mossformer2\cite{12-mossformer2-zhao2024mossformer2}, and speech separated by GSS. We assert that noisy speech preserves the original signal details, avoiding distortion; Mossformer2 focuses on low-quality audio and non-stationary noise, effectively separating the target speech; and the GSS-separated signal retains a more complete speech structure. The weighted fusion of these three signals facilitates the adaptive selection of the optimal signal combination.
The formula for the weighted fusion is as follows:

\begin{equation}
\widetilde{\boldsymbol{y}} = \omega_1 \text{Xsum} + \omega_2 \hat{\boldsymbol{y}} + \omega_3 \text{Xgss}
\end{equation}

where $\omega_1$, $\omega_2$, and $\omega_3$ represent the dynamic weighting coefficients that satisfy the condition $\omega_1 + \omega_2 + \omega_3 = 1$. $\text{Xsum}$ represents the multichannel summed signal with noise, $\hat{\boldsymbol{y}}$ represents the speech signal after Mossformer2 separation, and $\text{Xgss}$ represents the multichannel speech signal after GSS processing.

In the ASR system, the bridging module serves as a key component, primarily aimed at improving the quality of speech signals in noisy environments. Existing bridging modules typically rely on the SNR\cite{SNR-oa-chen2023noise} to predict the weighting coefficients for multiple signal paths. It first predicts the speech SNR using cosine similarity, then maps the OA coefficient through a linear layer, and is trained with mean squared error loss to ensure that the balanced speech approximates the clean speech. In contrast, we propose a sentence-level bridging module supervised by ASR recognition information, which predicts the OA coefficient from three input signals to generate an ASR-optimized input, as described below:

\begin{equation}
\widehat{\omega}=\text{Bridging}(\text{Xsum},\widehat{\boldsymbol{y}},\text{Xgss})
\end{equation}
Fbank features are selected as input for the bridging module and encoded using the Conformer architecture to extract high-level time-frequency features. Subsequently, convolutional and attention-based statistical pooling \cite{15-asp-okabe2018attentive,wang2024integrating} are applied to capture global contextual information and condense it into sentence-level feature representations. A fully connected layer then maps these representations to one-dimensional logits, which are normalized to the [0, 1] range using the Sigmoid activation function, ultimately producing the desired OA coefficient. The detailed process is illustrated in \autoref{fig:kunkun1}.

Since CER directly reflects ASR performance and aligns closely with the final objective, it serves as the supervisory criterion for training the bridging module. Specifically, the OA coefficient is determined based on ASR recognition information to balance the ratio of noisy speech, separated speech, and GSS-separated speech, thereby optimizing the training process. For each speech sample, CER values corresponding to different OA coefficients are precomputed to construct a vector $\boldsymbol{cers}$. Cosine similarity is then applied to ensure consistency between the predicted OA coefficient and the optimal CER distribution.

Specifically, for $\omega_1$, $(1/k + 2)$ fixed OA coefficients are pre-set with a step size of $k$ within the range $k \in [0, 0.05]$, with coefficients increasing from 0 to 1. For $\omega_2$, since performing a comprehensive grid search for both variables would substantially increase computational complexity, only 10 fixed discrete values are set with a step size of 0.1, ranging from 0 to 1. These coefficient pairs $(\omega_1, \omega_2)$ are then used in the SS, GSS, and ASR systems to process the training set. For each speech sample in the training set, the CER is calculated after processing with different OA coefficients, resulting in $10 \times (\frac{1}{k} + 2)$ sets of CER values. These WER values will guide the subsequent training of the bridging module, as described below:

% The speech recognition output guides the training of the bridging module. CER directly reflects ASR performance and is strongly correlated with the final objective, so we use CER as supervision for the bridging module training. Specifically, for $\omega_1$, we preset (1/k + 2) fixed OA coefficients with a step size of $k \in [0, 0.05]$, ranging from 0 to 1. For $\omega_2$, considering that performing a full grid search over two variables would significantly increase the complexity, we set 10 groups of fixed discrete values with a step size of 0.1, ranging from 0 to 1. These coefficient pairs $(\omega_1, \omega_2)$ are then used for SS, GSS, and ASR systems to process the training set. For each speech sample in the training set, after processing with different OA coefficients, the CER is calculated, resulting in $10 \times (\frac{1}{k} + 2)$ sets of CER values. These WER values are then used to guide the training of the bridging module, as shown below:

\begin{equation}
\mathcal{L}_{OA} = -\log \sigma \left(\frac{\boldsymbol{cs}(\boldsymbol{logits}, \sigma(\boldsymbol{cers}))}{\tau}\right)
\end{equation}
where $logits$ represents the unnormalized weights output by the bridging module, $\sigma$ denotes the sigmoid function, $cs$ refers to cosine similarity, $\boldsymbol{cers}$ represents the pre-computed CER vectors obtained through grid search, which reflect the ASR performance of different weight combinations, and $\tau$ represents the temperature coefficient smoothing distribution, which controls the smoothness of the distribution.

\section{Dataset}

\subsection{Speaker Diarization}
For the dereverberation model used in the diarization task, we trained it using near-field data from MISP2025 and generated room impulse responses using the Pyroomacoustics \cite{pyroomacoustics} library. Room dimensions were randomly set: length 3–6 meters, width 5–12 meters, and height fixed at 3 meters to simulate a realistic conference environment. The microphone position was randomly selected, with source-to-microphone distances of 1 or 3 meters. To align clean and reverberant speech, far-field audio was generated using absorption coefficients of 0.4 and 1.0. For the WavLM-based end-to-end model, far-field training data from Alimeeting, AISHELL4, MISP2021, and MISP2025 were used. The embedding extractors utilize the ResNet34 and ResNet293 models from the Wespeaker toolkit. In the traditional multi-module model, the pretrained SimAM-ResNet100 model from Wespeaker toolkit was used as the embedding extractor, with a PLDA model trained on the VoxCeleb2 \cite{voxceleb2} dataset.

\subsection{Speech Recognition}

For the ASR task, we utilized various data augmentation methods to generate approximately 1600 hours of training data, referred to as Multi-source Data Integration and Mixing Data Augmentation (MIM-DA). Specifically, MISP2025 training data, along with the open-source Alimeeting and AISHELL-4 datasets, as well as MISP2025-Near, Alimeeting-Near, and portions of the Kespeech dataset, were utilized to simulate approximately 1000 hours of 8-channel meeting scenario data\footnote[2]{https://github.com/jsalt2020-asrdiar/jsalt2020\_simulate}. The configuration followed a rectangular layout with 2 rows and 4 columns, with a microphone array width of 0.197 meters and a height of 0.134 meters\footnote[3]{https://github.com/jsalt2020-asrdiar/jsalt2020\_simulate/blob/master/\\configs/common/meeting\_reverb.json}. Additionally, the bridging module training data generation method from \autoref{subsec: asr} was applied to process the MISP2025 training set. Data augmentation techniques included speed perturbation, MUSAN noise addition, and dynamic SpecAugment. For multi-channel data, oracle RTTM labels were used to perform GSS, obtaining separated speech segments aligned with ASR annotations for ASR training. The simulated data was also used for training the Mossformer2 model.

\section{Experiments}
\subsection{Experimental Setup} 

For the end-to-end segmentation method, the clustering threshold was set to 0.65. For the traditional multi-module method, the weight coefficient of the dereverberation model's output audio was set to 0.7, and that of the original audio to 0.3, before performing the mixing operation. In the VBx clustering stage, the clustering threshold was set to 0, and the parameters were defined as $Fa$=0.15, $Fb$=5.5, and $loopP$=0.99.

In the SS front-end, GSS optimized overlapping speech using precomputed speaker activity and blind source separation, combined with WPE for dereverberation, CACGMM-based mask estimation, and MVDR beamforming to extract target speech. In the ASR back-end, we employed the high-accuracy, non-autoregressive Chinese recognition model Paraformer \cite{3-gao2022paraformer}, which used CIF prediction and MWER training. Additionally, we used SenseVoice-Small \cite{4-an2024funaudiollm}, an efficient multilingual model that featured an SAN-M encoder, task-specific embeddings, and CTC loss. The Conformer layer of bridging module was configured with 4 encoder layers, an output dimension of 256, and 4 attention heads, with a Conv2D input layer. During training, the OA coefficient step size was set to 0.05, the learning rate to 0.001, and the maximum number of training epochs to 30.

\subsection{Results}
\subsubsection{Speaker Diarization}

\autoref{tab:der_results} presented the DER (\%) for the audio-visual speaker diarization task. System S0 served as the baseline, achieving a DER of 15.52\% using audio features, multi-speaker lip ROIs, and i-vector embeddings as multi-modal inputs. System S1, employing an end-to-end segmentation method, achieved DER of 8.93\% on the development set and 10.62\% on the evaluation set. System S2 replaced the embedding extractor with ResNet293, significantly reduced DER on the development set but showed only slight improvement on the evaluation set. System S3 utilized a traditional multi-module diarization method, achieving DER of 8.69\% and 13.52\%, respectively, showing mixed performance compared to S1. Overlap speech inference in S1 suggested that traditional algorithms performed better with lower overlap proportions. System S4 incorporated dereverberation and mixing algorithms into the traditional pipeline, reducing DER from 8.69\% to 8.34\% on the development set. System S5 applied the "Overlap Decision" strategy to fuse S1 and S4, further reducing DER to 7.99\% and 9.09\% on the development and evaluation sets, confirming its effectiveness. Finally, System S6 fused S5 results from eight channels, achieving DER of 7.75\% and 8.88\%, representing a 42.78\% relative reduction from the baseline and ranking fourth in the AVSD track.

\begin{table}[htb]
    \centering
    \renewcommand{\arraystretch}{1.2}
    \caption{The DER (\%) for AVSD task using hybrid traditional and end-to-end segmentation methods.}
    \vspace{-0.5em} % 插入 1em 的垂直空白，可以根据需要调整
\begin{tabular}{cccc}
\toprule
\multirow{2}{*}{System} & \multirow{2}{*}{Method}    & \multicolumn{2}{c}{DER(\%)}   \\ \cline{3-4} 
                        &                            & DEV           & EVAL          \\ \hline
S0                      & Baseline                   & -             & 15.52         \\
S1                      & ResNet34 + WavLM E2E       & 8.93          & 10.62         \\
S2                      & ResNet293 + WavLM E2E      & 8.21          & 10.69         \\
S3                      & SimAM-ResNet100 + VBx      & 8.69          & 13.25         \\
S4                      & SimAM-ResNet100 + VBx + SE & 8.34          & -             \\
S5                      & Hybrid S1 + S4             & 7.99          & 9.09          \\
S6                      & S5 Dover-lap 8 channels    & \textbf{7.75} & \textbf{8.88} \\ \bottomrule
\end{tabular}
    \label{tab:der_results} % 给表格加标签，方便引用
    \vspace{-0.5em} % 插入 1em 的垂直空白，可以根据需要调整
\end{table}

% It is important to note that despite attempts to use video information to capture the activity of different speakers' lip regions, the diarization results based solely on audio could not be corrected when sufficient audio training data was available. Additionally, we found that the final error rate for the system S6 development set comprised 2.63$\%$ MISS errors, 4.52$\%$ False Alarm errors, and only 0.59$\%$ speaker errors. This suggests that most of the errors arise from VAD and overlap detection issues, and the higher False Alarm errors do not appear to significantly impact the ASR system. These errors occur because non-speech portions are mistakenly labeled as speech. Notably, the audio-only, single-channel results for system S5 on the evaluation set were only 1.00$\%$ worse than the top-ranked system in this track, suggesting that further exploration is required to refine the correction of audio-based diarization results using video modalities.

\vspace{-\baselineskip}
\subsubsection{Speech Recognition}

\autoref{tab:cer_results} presented the CER (\%) results for various combinations of front-end processing strategies and recognition models on the MISP2025 test set. The results indicated that, under the same front-end conditions, the Paraformer model consistently outperformed the SenseVoice model due to its architectural advantages. Specifically, when using GSS as the front-end, System A3 (Paraformer) achieved CER of 6.47\% and 11.24\% on the Dev and Eval sets, respectively, slightly outperforming System A1 (SenseVoice), which achieved CER of 6.51\% and 11.46\%. Similarly, with MIM-DA as the front-end, System A4 (Paraformer) achieved CER of 6.13\% and 10.63\% on the Dev and Eval sets, surpassing System A2 (SenseVoice), which recorded 6.40\% and 10.84\%. Additionally, we reproduced the SNR-based OA method from \cite{SNR-oa-chen2023noise} and found that it struggled to balance the weighting coefficients of noisy speech, SS-separated speech, and GSS-separated speech, leading to suboptimal performance in conference scenarios. In contrast, our proposed ASR-aware OA method benefited from ASR back-end supervision with real-world data, achieving the best single-system performance across all experiments, with CER of 5.91\% and 10.09\% on the Dev and Eval sets, respectively. Finally, by applying ROVER\cite{rover-fiscus1997post} to fuse all systems, CER further decreased to 5.54\% and 9.48\%, representing a 52.62\% reduction compared to the official baseline.

\begin{table}[htb]
    \centering
    \renewcommand{\arraystretch}{1.2}
    \caption{The CER ($\%$) results for various combinations of front-end processing strategies.}
    \vspace{-0.5em} % 插入 1em 的垂直空白，可以根据需要调整
 \begin{tabular}{ccccc}
\toprule
\multirow{2}{*}{System} & \multirow{2}{*}{Front-end} & \multirow{2}{*}{Model} & \multicolumn{2}{c}{CER (\%)}      \\ \cline{4-5} 
                        &                            &                        & DEV             & EVAL            \\ \hline
A0                      & \multicolumn{2}{c}{Baseline}                        & -               & 20.01         \\
A1                      & GSS                        & Sensevoice             & 6.51          & 11.46         \\
A2                      & MIM-DA                     & Sensevoice             & 6.40          & 10.84        \\
A3                      & GSS                        & Paraformer             & 6.47          & 11.24         \\
A4                      & MIM-DA                     & Paraformer             & 6.13          & 10.63         \\
A5                      & SNR OA                     & Paraformer             & 7.11          & -               \\
A6                      & ASR-Aware OA               & Paraformer             & 5.91          & 10.09         \\
A7                      & \multicolumn{2}{c}{ROVER}                           & \textbf{5.54} & \textbf{9.48} \\ \bottomrule
\end{tabular}
    \label{tab:cer_results} % 给表格加标签，方便引用
\end{table}

% When attempting to integrate the video modality into the recognition tasks, we found that the lip features extracted by the front-end model were insufficiently effective, leading to significant degradation in the quality of the visual information. This low-quality visual modality not only failed to supplement the audio information effectively, but also introduced interfering noise during the fusion process, undermining the performance of the dominant audio signal.
\vspace{-\baselineskip}
\subsubsection{Diarization and Recognition}

\autoref{tab:final_results} presented the cpCER (\%) for the AVSR track after combining the methods proposed in the first two tracks. The RTTM files output by the diarization system were passed into the GSS and ASR module for processing. System M0 presented the performance of the baseline method on the test set, achieving a cpCER of 84.05\%. By combining the S1 system from Track 1 with the A3 system from Track 2, the cpCER was reduced to 13.57\%. By combining the S6 system from Track 1 with the A6 system from Track 2, the cpCER was further reduced to 11.56\%, representing an 86.25\% reduction relative to the baseline. Our final submission ranked first in Track 3.

\begin{table}[htb]
    \centering
    \renewcommand{\arraystretch}{1.2}
    \caption{The cpCER (\%) for the AVSR track after combining the methods proposed in the first two tracks.}
    \vspace{-0.5em} % 插入 1em 的垂直空白，可以根据需要调整
\begin{tabular}{ccc}
\toprule
System & Method              & EVAL cpCER(\%) \\ \hline
M0     & Baseline            & 84.05          \\
M1     & S1 + A3 & 13.57          \\
M2     & S6 + A6 & \textbf{11.56} \\ \bottomrule
\end{tabular}
    \label{tab:final_results} % 给表格加标签，方便引用
\end{table}

\vspace{-\baselineskip}

\section{Conclusion}

This paper described our diarization and ASR systems for the MISP 2025 Challenge. For the diarization task, we combined traditional multi-module methods with a WavLM-based end-to-end segmentation model. For the ASR task, the proposed ASR-Aware OA method effectively addressed speech enhancement challenges in low-SNR environments, achieving a CER of 9.48\% on the evaluation set. By integrating the diarization and ASR systems, we ultimately achieved a cpCER of 11.56\% in the combined task. We secured first place in both Track 2 and Track 3. Although we explored the use of video information for both diarization and ASR tasks, the final results did not demonstrate significant improvements. We found that real-world challenges, such as blurred lip images and poor lighting, often resulted in extracted visual features that interfered with, rather than assisted, the audio system. In the future, we will further explore methods to effectively utilize video information to overcome the performance bottleneck of audio-only systems, particularly in scenarios where audio quantity and quality dominate.

\section{Acknowledgements}
This work is supported by National Engineering Research Center of Multi-dimensional Identification and Trusted Authentication Technology (No.IDNERC202404).

\bibliographystyle{IEEEtran}
\bibliography{mybib}

\begin{thebibliography}{10}
\providecommand{\url}[1]{#1}
\csname url@samestyle\endcsname
\providecommand{\newblock}{\relax}
\providecommand{\bibinfo}[2]{#2}
\providecommand{\BIBentrySTDinterwordspacing}{\spaceskip=0pt\relax}
\providecommand{\BIBentryALTinterwordstretchfactor}{4}
\providecommand{\BIBentryALTinterwordspacing}{\spaceskip=\fontdimen2\font plus
\BIBentryALTinterwordstretchfactor\fontdimen3\font minus \fontdimen4\font\relax}
\providecommand{\BIBforeignlanguage}[2]{{%
\expandafter\ifx\csname l@#1\endcsname\relax
\typeout{** WARNING: IEEEtran.bst: No hyphenation pattern has been}%
\typeout{** loaded for the language `#1'. Using the pattern for}%
\typeout{** the default language instead.}%
\else
\language=\csname l@#1\endcsname
\fi
#2}}
\providecommand{\BIBdecl}{\relax}
\BIBdecl

\bibitem{misp2025zongjie}
\BIBentryALTinterwordspacing
M.~Gao, S.~Wu, H.~Chen, J.~Du, C.-H. Lee, S.~Watanabe, J.~Chen, S.~S. Marco, and O.~Scharenborg, ``The multimodal information based speech processing (misp) 2025 challenge: Audio-visual diarization and recognition,'' 2025. [Online]. Available: \url{https://arxiv.org/abs/2505.13971}
\BIBentrySTDinterwordspacing

\bibitem{2025misp}
H.~Chen, C.-H.~H. Yang, J.-C. Gu, S.~M. Siniscalchi, and J.~Du, ``{MISP-Meeting}: A real-world dataset with multimodal cues for long-form meeting transcription and summarization,'' in \emph{Proceedings of the 63st Annual Meeting of the Association for Computational Linguistics (Volume 1: Long Papers)}.\hskip 1em plus 0.5em minus 0.4em\relax Association for Computational Linguistics, 2025, pp. 1--14.

\bibitem{introduction1}
C.~B. Boeddeker, T.~Cord-Landwehr, T.~v. Neumann, and R.~Haeb-Umbach, ``Multi-stage diarization refinement for the chime-7 dasr scenario,'' in \emph{Proc. CHiME 2023}, 2023, pp. 51--56.

\bibitem{introduction2}
N.~Kamo, N.~Tawara, A.~Ando, T.~Kano, H.~Sato, R.~Ikeshita, T.~Moriya, S.~Horiguchi, K.~Matsuura, A.~Ogawa \emph{et~al.}, ``Ntt multi-speaker asr system for the dasr task of chime-8 challenge,'' \emph{arXiv preprint arXiv:2409.05554}, 2024.

\bibitem{introduction3}
M.~Cheng, H.~Wang, Z.~Wang, Q.~Fu, and M.~Li, ``The whu-alibaba audio-visual speaker diarization system for the misp 2022 challenge,'' in \emph{ICASSP 2023-2023 IEEE International Conference on Acoustics, Speech and Signal Processing (ICASSP)}.\hskip 1em plus 0.5em minus 0.4em\relax IEEE, 2023, pp. 1--2.

\bibitem{introduction4}
T.~Liu, Z.~Chen, Y.~Qian, and K.~Yu, ``Multi-speaker end-to-end multi-modal speaker diarization system for the misp 2022 challenge,'' in \emph{ICASSP 2023-2023 IEEE International Conference on Acoustics, Speech and Signal Processing (ICASSP)}.\hskip 1em plus 0.5em minus 0.4em\relax IEEE, 2023, pp. 1--2.

\bibitem{9-misp-3wang2023multimodal}
Z.~Wang, S.~Wu, H.~Chen, M.-K. He, J.~Du, C.-H. Lee, J.~Chen, S.~Watanabe, S.~Siniscalchi, O.~Scharenborg \emph{et~al.}, ``The multimodal information based speech processing (misp) 2022 challenge: Audio-visual diarization and recognition,'' in \emph{ICASSP 2023-2023 IEEE International Conference on Acoustics, Speech and Signal Processing (ICASSP)}.\hskip 1em plus 0.5em minus 0.4em\relax IEEE, 2023, pp. 1--5.

\bibitem{introduction6}
L.~Zhang, H.~Zhao, Y.~Li, B.~Pang, Y.~Wang, H.~Wang, W.~Rao, Q.~Wang, and L.~Xie, ``The flyspeech audio-visual speaker diarization system for misp challenge 2022,'' \emph{arXiv preprint arXiv:2307.15400}, 2023.

\bibitem{introduction7}
N.~Tawara, M.~Delcroix, A.~Ando, and A.~Ogawa, ``Ntt speaker diarization system for chime-7: multi-domain, multi-microphone end-to-end and vector clustering diarization,'' in \emph{ICASSP 2024-2024 IEEE International Conference on Acoustics, Speech and Signal Processing (ICASSP)}.\hskip 1em plus 0.5em minus 0.4em\relax IEEE, 2024, pp. 11\,281--11\,285.

\bibitem{introduction8}
S.~Huang, Y.~Du, Y.~Wang, J.~Deng, and R.~Zheng, ``The fosafer system for the icassp2024 in-car multi-channel automatic speech recognition challenge,'' in \emph{2024 IEEE International Conference on Acoustics, Speech, and Signal Processing Workshops (ICASSPW)}.\hskip 1em plus 0.5em minus 0.4em\relax IEEE, 2024, pp. 5--6.

\bibitem{wavlm}
J.~Han, F.~Landini, J.~Rohdin, A.~Silnova, M.~Diez, and L.~Burget, ``Leveraging self-supervised learning for speaker diarization,'' \emph{arXiv preprint arXiv:2409.09408}, 2024.

\bibitem{pyannote3}
A.~Plaquet and H.~Bredin, ``Powerset multi-class cross entropy loss for neural speaker diarization,'' \emph{Proc. Interspeech}, pp. 3222--3226, 2023.

\bibitem{mp}
Y.-X. Lu, Y.~Ai, and Z.-H. Ling, ``{MP-SENet}: A speech enhancement model with parallel denoising of magnitude and phase spectra,'' in \emph{Proc. Interspeech}, 2023, pp. 3834--3838.

\bibitem{simam}
L.~Yang, R.-Y. Zhang, L.~Li, and X.~Xie, ``Simam: A simple, parameter-free attention module for convolutional neural networks,'' in \emph{International conference on machine learning}.\hskip 1em plus 0.5em minus 0.4em\relax PMLR, 2021, pp. 11\,863--11\,874.

\bibitem{wespeaker}
H.~Wang, C.~Liang, S.~Wang, Z.~Chen, B.~Zhang, X.~Xiang, Y.~Deng, and Y.~Qian, ``Wespeaker: A research and production oriented speaker embedding learning toolkit,'' in \emph{ICASSP 2023-2023 IEEE International Conference on Acoustics, Speech and Signal Processing (ICASSP)}.\hskip 1em plus 0.5em minus 0.4em\relax IEEE, 2023, pp. 1--5.

\bibitem{vbx}
F.~Landini, J.~Profant, M.~Diez, and L.~Burget, ``Bayesian hmm clustering of x-vector sequences (vbx) in speaker diarization: theory, implementation and analysis on standard tasks,'' \emph{Computer Speech \& Language}, vol.~71, p. 101254, 2022.

\bibitem{dover}
D.~Raj, L.~P. Garcia-Perera, Z.~Huang, S.~Watanabe, D.~Povey, A.~Stolcke, and S.~Khudanpur, ``Dover-lap: A method for combining overlap-aware diarization outputs,'' in \emph{2021 IEEE Spoken Language Technology Workshop (SLT)}.\hskip 1em plus 0.5em minus 0.4em\relax IEEE, 2021, pp. 881--888.

\bibitem{11-gss-raj2022gpu}
D.~Raj, D.~Povey, and S.~Khudanpur, ``Gpu-accelerated guided source separation for meeting transcription,'' \emph{arXiv preprint arXiv:2212.05271}, 2022.

\bibitem{14-han2024audio}
R.~Han, X.~Yan, W.~Xu, P.~Guo, J.~Sun, H.~Wang, Q.~Lu, N.~Jiang, and L.~Xie, ``An audio-quality-based multi-strategy approach for target speaker extraction in the misp 2023 challenge,'' \emph{arXiv preprint arXiv:2401.03697}, 2024.

\bibitem{1-huang2024fosafer}
S.~Huang, D.~Zhang, Y.~Wang, J.~Deng, and R.~Zheng, ``The fosafer system for the chime-8 mmcsg challenge,'' in \emph{CHiME Workshop on Speech Processing in Everyday Environments}, 2024.

\bibitem{2-huang2024enhanced}
S.~Huang, D.~Zhang, J.~Deng, and R.~Zheng, ``Enhanced asr for stuttering speech: Combining adversarial and signal-based data augmentation,'' in \emph{2024 IEEE Spoken Language Technology Workshop (SLT)}.\hskip 1em plus 0.5em minus 0.4em\relax IEEE, 2024, pp. 393--400.

\bibitem{5-oa-1cui2025reducing}
Z.~Cui, C.~Cui, T.~Wang, M.~He, H.~Shi, M.~Ge, C.~Gong, L.~Wang, and J.~Dang, ``Reducing the gap between pretrained speech enhancement and recognition models using a real speech-trained bridging module,'' \emph{arXiv preprint arXiv:2501.02452}, 2025.

\bibitem{12-mossformer2-zhao2024mossformer2}
S.~Zhao, Y.~Ma, C.~Ni, C.~Zhang, H.~Wang, T.~H. Nguyen, K.~Zhou, J.~Q. Yip, D.~Ng, and B.~Ma, ``Mossformer2: Combining transformer and rnn-free recurrent network for enhanced time-domain monaural speech separation,'' in \emph{ICASSP 2024-2024 IEEE International Conference on Acoustics, Speech and Signal Processing (ICASSP)}.\hskip 1em plus 0.5em minus 0.4em\relax IEEE, 2024, pp. 10\,356--10\,360.

\bibitem{SNR-oa-chen2023noise}
Y.-W. Chen, J.~Hirschberg, and Y.~Tsao, ``Noise robust speech emotion recognition with signal-to-noise ratio adapting speech enhancement,'' \emph{arXiv preprint arXiv:2309.01164}, 2023.

\bibitem{15-asp-okabe2018attentive}
K.~Okabe, T.~Koshinaka, and K.~Shinoda, ``Attentive statistics pooling for deep speaker embedding,'' \emph{arXiv preprint arXiv:1803.10963}, 2018.

\bibitem{wang2024integrating}
Y.~Wang, Y.~Du, D.~Zhang, R.~Zheng, and J.~Deng, ``Integrating self-supervised pre-training with adversarial learning for synthesized song detection,'' in \emph{2024 IEEE Spoken Language Technology Workshop (SLT)}.\hskip 1em plus 0.5em minus 0.4em\relax IEEE, 2024, pp. 795--802.

\bibitem{pyroomacoustics}
R.~Scheibler, E.~Bezzam, and I.~Dokmani{\'c}, ``Pyroomacoustics: A python package for audio room simulation and array processing algorithms,'' in \emph{2018 IEEE international conference on acoustics, speech and signal processing (ICASSP)}.\hskip 1em plus 0.5em minus 0.4em\relax IEEE, 2018, pp. 351--355.

\bibitem{voxceleb2}
J.~S. Chung, A.~Nagrani, and A.~Zisserman, ``Voxceleb2: Deep speaker recognition,'' \emph{arXiv preprint arXiv:1806.05622}, 2018.

\bibitem{3-gao2022paraformer}
Z.~Gao, S.~Zhang, I.~McLoughlin, and Z.~Yan, ``Paraformer: Fast and accurate parallel transformer for non-autoregressive end-to-end speech recognition,'' \emph{arXiv preprint arXiv:2206.08317}, 2022.

\bibitem{4-an2024funaudiollm}
K.~An, Q.~Chen, C.~Deng, Z.~Du, C.~Gao, Z.~Gao, Y.~Gu, T.~He, H.~Hu, K.~Hu \emph{et~al.}, ``Funaudiollm: Voice understanding and generation foundation models for natural interaction between humans and llms,'' \emph{arXiv preprint arXiv:2407.04051}, 2024.

\bibitem{rover-fiscus1997post}
J.~G. Fiscus, ``A post-processing system to yield reduced word error rates: Recognizer output voting error reduction (rover),'' in \emph{1997 IEEE Workshop on Automatic Speech Recognition and Understanding Proceedings}.\hskip 1em plus 0.5em minus 0.4em\relax IEEE, 1997, pp. 347--354.

\end{thebibliography}

\end{document}